\documentclass[12pt,dvips]{iopart}
\usepackage{iopams}
\usepackage{graphicx}
\usepackage{epsfig,graphicx}
\usepackage{xcolor}

\begin{document}

\title{\bf Noise and nonlinearities in high-throughput data.}
\author{Viet-Anh Nguyen$^{(1)}$, Zdena Koukol\'{i}kov\'{a}-Nicola$^{(2)}$, Franco Bagnoli$^{(3)}$, Pietro Li\'o$^{(1)}$}
\address{$^1$ Computer Laboratory, University of Cambridge, Cambridge CB3 0FD, UK}
\address{$^2$ Fachhochschule Nordwestschweiz, Hochschule fr Technik, Steinackerstrasse 5, CH-5210 Windisch, Switzerland}
\address{$^3$ Department of Energy, University of Florence,
 S. Marta, 3 50139 Firenze. \\Also CSDC and INFN, sez. Firenze.}
\ead{franco.bagnoli@unifi.it}

\begin{abstract}
High throughput data analysis are becoming common in biology, communications, economics and sociology. This vast amount of data are usually represented in form of
matrices and can be considered as knowledge networks. Spectral-based approaches have been proved useful in extracting
hidden information within such networks and to estimate missing data, but these methods are based essentially on linear assumptions. The physical models of matching, when available, often suggest  nonlinear mechanisms, that may sometimes be identified as noise. The use of nonlinear models in data analysis, however, may require the introduction of many parameters, that lowers the statistical weight of the model. According with the quality of data, a simpler linear analysis may be more convenient than more complex approaches.

In this paper, we show how a simple nonparametric Bayesian model may be used to explore the role of nonlinearities and noise in synthetic and experimental datasets.
\end{abstract}
\noindent{\it Keywords\/}: Special issue, Data mining (Theory)

\section{Introduction}
There is currently a tremendous growth in the amount of high throughput data extracted from life sciences, electronic communications, economics and social sciences.
Examples of high throughput life science data are the large number of completely sequenced genomes, 3D protein structures, DNA chips, and mass spectroscopy data. Large amounts of data are distributed across many sources over the web, with high degree of semantic heterogeneity and different levels of quality.
These data must be combined with other data and processed by statistical tools for patterns, similarities, and unusual occurrences to be observed.

The results of many experiments can be summarized in a large
matrix, where rows represent repetition of the experiment in
different contexts, and the columns are the output of a single
measurement. Let us consider the following cases:
microarray sampling, protein-substrate affinity, socio-psychological surveys.
Let us illustrate the similarities of these examples.

\subsection{Examples of datasets}

\subsubsection{Microarray data}

A DNA microarray (gene chip) can be seen as an ordered collection of
spots, on each of which there is a different probe formed by
known sequences of cDNA. A sample of mRNA, supposed to represent the
gene expressed in a given  tissue under investigation, is let
hybridize with the probes. Fluorescent techniques allows to detect the
hybridized spots. The idea is that of using probes specific for a
unique region of a gene, detecting the genes expressed in a
tissue. The experiment is repeated for many tissue, from different
parts of the body, from different patients, or from a different
phase of the cellular cycle. The data can therefore be arranged using
the probe numbering as column index, and tissue numbering as row
index. The goal is that of identifying the difference in gene
expressions in the different cases.

There are many problems in extracting information from these data.
Some data may be missing, control spots are sometimes more hybridized
that normal ones, low-intensity data cannot be easily distinguished
from noise.

\subsubsection{Protein-substrate affinity}

A similar problem is that of investigating the shape of a protein or
of a peptide. The interaction of proteins with the outer world (in
particular concerning the immune response) depends on the shape.
At present, it is not possible to reconstruct the
tri-dimensional shape of a protein from its primary sequence (easily
obtained by mRNA sequence). Moreover, proteins very often
glycosylated, and these sugar chains attached to the outer
surface may be the most important factor for inflammation.
On the other hand, direct visualization of protein surface, using
NMR, electronic microscopy, etc. is a very slow and costly process.

A method for obtaining information about this shape is that of using
protein or antibody arrays, similar to DNA microarrays. Again, in
this case, the pattern of matches can be represented as a matrix,
with columns corresponding to substrates (probing proteins or
antibodies) and rows to different proteins under investigation.

\subsubsection{Questionnaires and other socio-psychological data}

The high-level investigation of the human mind take often the form of
the study of responses to stimuli. The stimulus may be
planned and targeted, like in the case of
questionnaires, or occasional/unplanned like for instance those that lead to
choosing some good. In this case, it is economically advantageous to study the
 patterns that emerge, for instance in renting  DVDs~\cite{netflix}, opinions on books~\cite{amazon}, supermarket tickets. Also Google page rank~\cite{google} may be considered in this class; in this case the ``opinions'' are the links that bring to the page under investigation.

All these data may be (ideally) represented in matrix form, with
rows corresponding to customers, and columns corresponding to items
or goods.

In this case, in addition to the usual problems of consistency and
noise, there is a special meaning in missing data: an accurate method
for ``anticipating'' them from the knowledge stored in the matrix
would constitute a valuable tool for personal
advertising.

However, in the case of humans, one should consider also that
tastes change and evolve in time.

\subsection{Knowledge networks}

The extraction of  information about the properties of the gene,
proteins and humans is performed using statistical tools,
mainly based on variations of singular value
decomposition~\cite{troya01}. The goal is that of extracting the most
robust characteristics of patterns, clustering the data in similarity
classes, reconstruct missing data, detect outliers, reduce
noise. it is rather unusual to take into consideration an
explicit model for the generation of data, \emph{i.e.}, for the
matching mechanism.

The problem may be reformulated in geometrical terms. We shall denote
with the word ``probe'' the substrates or the questionnaires, and
with the word ``subject'' the mRNAs, the proteins and the individuals
of the three examples. Actually, the role of probes and subjects are symmetric, the only difference is that i sgeneral one has more ``a priori'' knowledge of probes that of subjects. 

A subject can be visualized as an array of $M$ ``tastes'', and the probes as a complementary array of ``characteristics''. In the case of mRNA, this space is just the sequence space of basis, in case of proteins it is a way of coding the surface (motifs), in case of psychological data, these are mental modules, often called factorial ``dimensions''. The match between tastes and characteristics is denoted ``opinion''.

The match between tastes and characteristics may be linear, like a
scalar product, or highly nonlinear, as in the case of
protein-antibody or microarray interactions. In case of one-or-none interaction,
there is no noise and no inference can be performed on missing data.

In the linear case, the results are much more blurred, there is a
non-zero overlap between different samples, and it has been
shown~\cite{maslov01} that, if one knows a sufficient number of overlaps
between subjects, there is a rigid percolation threshold that in
principle allows the reconstruction of any ``taste'' once that one is
known. However, tastes are in general hidden or difficult to be
obtained. If one has at his disposition a sufficient amount of data,
it can be shown~\cite{bagnoli04} that the correlation between
expressed opinions approximates the real overlap among tastes. This
would in principle allow the perfect reconstruction of missing opinions
and detection of outliers. 

\subsection{Paper outline}

In this paper we investigate the role of nonlinearities and noise in
the matching phase. In particular, it is shown that nonlinearities
appear as noise when linear investigation tools are used. We
investigate the influence of nonlinearities in the rigid percolation
transition. Our recent proceeding paper \cite{nguyen08} could be used
as a complementary source, from which we suggest the use of correlation distribution among
subjects to predict the difference between random noise and nonlinearity.

The goal of this paper  is to explore
the limits of linear prediction, coupled with Bayesian tuning of
parameters, in predicting the features of linear and nonlinear systems, in the presence of an
eventual noise. As expected, the reconstruction of missing values works quite
well with the linear model, since it actually consists in a linear
interpolation. However, it works also for nonlinear models. Indeed, sometimes
we have information about the real physical process of matching, like in the
case of microarrays or proteins, while in other cases (opinions) this information
is unknown. So, let us concentrate on the more ``favorable'' case. Even in this
case, we are not sure to have taken into consideration all possible sources of
nonlinearities. Since it is rather
difficult to develop a complete matching model, one has to be content with a
"phenomenological" one, with a certain number of phenomenological parameters to
be tuned. As usual, the preference of a more complex model, with more
parameters, with respect to a simpler one should be justified on the basis of
the quality of available data. Therefore, one has to start with the simplest
model, usually a linear one, and tune the parameters (done here by Bayesian
estimators). Given the amount of ``explained'' signal by this model, one can
decide if it is worth to use a more complex model. In this paper we show
that even for synthetic data, where the matching model is perfectly controlled,
most of the information in the signal may be captured by linear methods also in the
presence of nonlinearities. In a future work we shall explore the properties of
nonlinear analysis.

We apply a recursive technique~\cite{maslov01, bagnoli04} to synthetic
data, obtained from linear matching models, and investigate the
limits of the ``black box'' reconstruction model. We also
investigate the role of nonlinearities and noise in the matching phase. In particular,
it is shown that nonlinearities appear as noise when linear investigation tools are used. We propose the use of Bayesian statistics to automatically determine the most appropriate value at each iteration of the learning process. As the result, the value of $M$ is changed accordingly to the amount of information available at each step, and approaches a fixed value when the predictions start to converge.

\section{The model}

Let us denote by $N$ the number of subjets, and by $D$ the number of probes or substrates. We assume that subject $i$ is represented in a (hidden) $M$-dimensional space as a vector $\mathbf{x}_i=(x_i^{(1)}, x_i^{(2)}, \dots,x_i^{(M)})$. The quantities $x_i^{(k)}$ can in principle be arbitrary, but we can assume that they are  normalized in the interval $[-1,1]$. A substrate $j$ is represented similarly by a vector $\mathbf{w}_j=\{w_j^{(k)}\}$ in a dual space, with $j=1, \dots, D$ and $k=1, \dots, M$.

The match $y_{ij}$ between a subject and a substrate is supposed to be a function of a the concordance between the characteristics
\[
  y_{ij} = f\left(\sum_k a^{(k)}  w_j^{(k)} x_i^{(k)} \right),
\]
where $a^{(k)}$ denotes the weight assigned to component $k$ (this weight could be eliminated by using non-normalized characteristics). For simplicity in the following we assume $a^{(k)}=1/M$, so that the argument of the function $f$ is also normalized.

The match $y_{ij}$ is the output of the experiment, and is the subject of data analysis in order to extract hidden information. For instance, one would like to know if data are consistent, how to reconstruct missing entries, how to use data for clustering subjects and substrates in terms of similarities, how to choose the optimal set of subjects for a given classification task.

The problems in data analysis arises from the role of noise and nonlinearities. We study three topical cases:
\begin{enumerate}
 \item linear case: $f(x)=x$, with no noise added. This is a reference case, and corresponds to the maximum of information that can be extracted.
 \item noisy case: $f(x) = x+\epsilon$, where $\epsilon$ is some gaussian noise. This is the model that is tacitly assumed in most of data analysis.
 \item nonlinear case: $f(x)$ is some nonlinear function, for instance $f(x)=\tanh(\beta x)$, where $\beta$ controls the influence of nonlinearities. This particular choice of the matching function corresponds to a case of \emph{thresholded} match, but other choices could be more suited, see for instance~\cite{DiPatti}.
\item nonlinear case with noise: a combination of case 2 and 3, which is presumably the nearer to reality.
\end{enumerate}

For convenience, we rewrite the standard assumption (case 2) in matrical form
\begin{equation} \label{trans_eq}
\mathbf{y}_{i} = \mathbf{W}\mathbf{x}_{i} + \boldsymbol{\epsilon}.
\end{equation}

One of the first target problem is the consistency check, or reconstruction of missing items. Suppose that some data $y_{i^*j^*}$ is missing. Can it be reconstructed from the rest of data? This should be possible if the number of subjects and substrates is large enough, and if their characteristics cover evenly enough the available space. On the contrary, if the missing data correspond to a particular match, not present in all other data, recovery is impossible. For instance, let us assume that only subject $i^*$ has characteristic $k^*$ different from zero, and has average values for all other characteristics. Therefore, this subject is the only sampling the ``dimension'' $k^*$, and its corresponding entry contains information not present in the rest of the database. On the contrary, if individual $i^*$ is exactly the same of another individual, this similarity should emerge (except for noise), and allow the perfect reconstruction.



The problem of reconstruction can therefore serve as a consistency check, and also as a tool for pointing out the data that deserve further investigations, either because they are spurious, or because they contain ``original'' information.

\section{Inference on missing values}

\subsection{Estimating the length of (hidden) feature space}
Looking at the model in (\ref{trans_eq}) from the regression view, we
consider $\mathbf{y}$ as the observed variables from the experimental space, $\mathbf{x}$ as
the hidden variables from the feature space, $\mathbf{W}$ as the model parameter that
relates these two sets of variables, and error $\boldsymbol{\epsilon}$ to follow the Gaussian distribution $\mathcal{N}(0,\sigma^2\mathbf{I})$.

We assume $\mathbf{x}_{i}$ to follow the standard Gaussian distribution
$\mathcal{N}(0,\mathbf{I})$. The diagonal unit variance implies that all vector components
are independent, which is reasonable enough. By integrating out $\mathbf{x}$, we get the
likelihood distribution of the data given all parameters:
\begin{equation}
p(\mathbf{Y}|\mathbf{W},\sigma^2) = (2\pi)^{-ND/2}
|\mathbf{V}|^{-N/2}
\exp\left(-\frac{1}{2}\tr\left((\mathbf{V})^{-1}\mathbf{P}\right)\right),
\end{equation}
where $\mathbf{V}=\mathbf{W}\mathbf{W}^{T}+\sigma^2\mathbf{I}$ and $\mathbf{P}=\mathbf{Y}\mathbf{Y}^{T}$.

Tipping and Bishop \cite{tip99} has used maximum likelihood to estimate $\mathbf{W}$ as:
\begin{equation} \label{w_est_eq}
\mathbf{\tilde{W}} = \mathbf{U}_{M}(\mathbf{\Lambda}_{M} - \sigma^2\mathbf{I})^{1/2}
\end{equation}
where the ${N}\times{M}$ matrix $\mathbf{U}_{M}$ is constructed by $M$ principal eigenvectors of
$\mathbf{Y}\mathbf{Y}^{T}$, the ${M}\times{M}$ matrix $\mathbf{\Lambda}_{M}$ contains $M$ largest
eigenvalues of $\mathbf{Y}\mathbf{Y}^{T}$. The arbitrary rotation matrix as in \cite{tip99} was
effectively selected as $\mathbf{I}$ for simplicity. The square root operation is safe with the
corresponding estimation of $\sigma^2$.

However, $M$ is normally not a known and fixed property in real systems. Even in the case in which we have information about the matching mechanism, it is more correct to treat $M$ as an unknown quantity, especially in the case in which we exploit linear correlations to obtain information about a nonlinear matching model. Maslov and Zhang \cite{maslov01}
has proposed a conjecture to effectively estimate $M$ using the knowledge of portion of
missing values of symmetric data matrix. Although the conjecture sometimes prove useful \cite{kouko08},
its usability limits to the case of symmetric input data. Here, we propose the use of Bayesian
approach to estimate $M$ from its posterior distribution. In other words, we want to calculate:
\begin{equation} \label{bayesM_eq}
p(M|\mathbf{Y}) = \frac{p(\mathbf{Y}|M)p(M)}{\int p(\mathbf{Y}|M)p(M)dM}
\end{equation}
where the likelihood of the data given $M$ is computed by integrating over all unknown parameters:
\begin{equation}
p(\mathbf{Y}|M) = \int_{\mathbf{X},\mathbf{W},{\sigma^2}}
p(\mathbf{Y}|\mathbf{W},\sigma^2)p(\mathbf{W}|\sigma^2,M)p(\sigma^2|M)
d\mathbf{W}d\sigma^2
\end{equation}

There is no closed-form solution to the model. A few papers \cite{everson00,rajan97,minka00} proposing different ways to estimate the sufficient number of principal
components to keep when performing PCA, which is very close in nature to our problem. To keep the
lightweight characteristic of the spectral algorithm, we base our calculation on the Laplace approximation of
$p(\mathbf{Y}|M)$ proposed by \cite{minka00}, which leads to the following estimation:
\begin{equation} \label{likeM_eq}
p(\mathbf{Y}|M) \approx N^{-\frac{M}{2}}(2\pi)^{\frac{(D-M+1)M}{2}}
\left(\prod_{i=1}^M{\lambda}_{i}\right)^{-\frac{N}{2}}
\left(\frac{\sum_{i=M+1}^D{\lambda}_{i}}{D-M}\right)^{-\frac{N(D-M)}{2}}
|\mathbf{A}|^{-\frac{1}{2}}
\end{equation}
where $|\mathbf{A}| = \prod_{i=1}^{M}\prod_{j=i+1}^{D}N
({\lambda}_{j}^{-1}-{\lambda}_{i}^{-1})({\lambda}_{i}-{\lambda}_{j})$
and ${\lambda}_{i}, i=1..D$ are the square root of the eigenvalues of $\mathbf{Y}\mathbf{Y}^{T}$.

Maslov and Zhang \cite{maslov01} provided an estimation of ${M}_{\mbox{\tiny eff}}$ - the sufficient number of
eigenvalues to keep during matrix reconstruction, taking into account the
proportion of missing values of the data matrix. Here we adapt this conjecture to
the case of asymmetric data, and use it as the suggestion for the upper boundary
of $M$ given $m$ missing elements. We define the prior distribution $p(M)$ as:
\begin{equation} \label{priorM_eq}
p(M) = \left\{ \begin{array}{lcl}
\frac{k}{(k-1){M}_{\mbox{\tiny eff}}+D} & \mbox{\tiny for} & M<=M_{\mbox{\tiny eff}}
\\ \frac{1}{(k-1){M}_{\mbox{\tiny eff}}+D} & \mbox{\tiny for} & M_{\mbox{\tiny eff}}<M<=D
\\ 0 & \mbox{\tiny otherwise}
\end{array}\right.
\end{equation}
where ${M}_{\mbox{\tiny eff}}=\frac{ND-m}{N}$ and the empirical value $k=3$.

\subsection{The algorithm}
Our proposed approximation algorithm  to infer on missing values of
data matrix $\mathbf{Y}$ is as follows:
\\(1) Construct the initial estimation $\mathbf{\tilde{Y}}$ of $\mathbf{Y}$ by assigning 0 to all unknown positions.
\\(2) Estimate the sufficient $\tilde{M}$ for $\mathbf{\tilde{Y}}$ using equations (\ref{bayesM_eq}), (\ref{likeM_eq}), and (\ref{priorM_eq}), the denominator in (\ref{bayesM_eq}) is ignored since it is a constant to $M$.
\\(3) Perform SVD on $\mathbf{\tilde{Y}}$, and construct the matrix $\mathbf{\tilde{Y}}^{'}$ by keeping the $\tilde{M}$ largest singular values and corresponding eigenvectors.
\\(4) Reconstruct $\mathbf{\tilde{Y}}$ from $\mathbf{\tilde{Y}}^{'}$ by filling known positions with their original values.
\\(5) Go to step (2).
\\We repeat this process until either there is no significant change on estimated values or a maximum number of iterations has been reached. 

The goal of the algorithm is to find a rank-M matrix that best approximates
the data matrix, or mathematically we want to find a solution for the following problem:
\[
  \min_{\mathbf{U}_{M},\mathbf{\Lambda}_{M},\mathbf{V}_{M}}{\parallel \mathbf{Y} - \mathbf{U}_{M}\mathbf{\Lambda}_{M}\mathbf{V}_{M}^{T} \parallel}^2
\]
When there is no missing values, the solution is simply the SVD of the complete data matrix.
For our iterative algorithm of reconstructing missing positions, the solution to this problem
will not change once it is found. More specifically, if we are given the solution and use it
to fill in the missing positions, the SVD of the resulted matrix will be exactly the given
solution. Hence, it is reasonable to believe that the algorithm will converge to the solution.
Our numerical experiments actually achieved very good convergence rate under various 
proportions of missing data.

\section{Experiments and Discussion}
\subsection{Synthetic Data}
\begin{figure}
  \centering
  (a)\includegraphics[width=6cm]{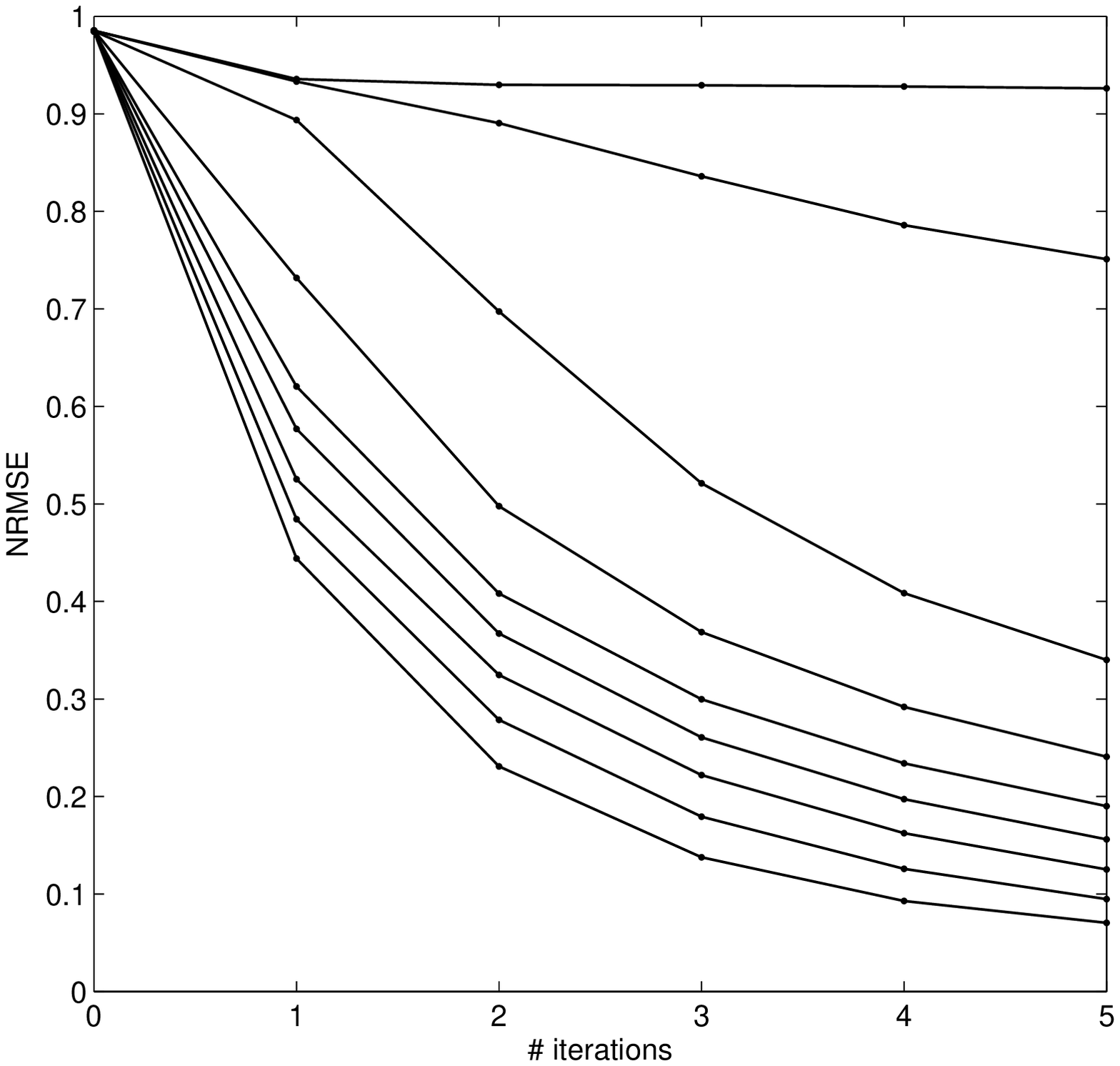}
  (b)\includegraphics[width=6cm]{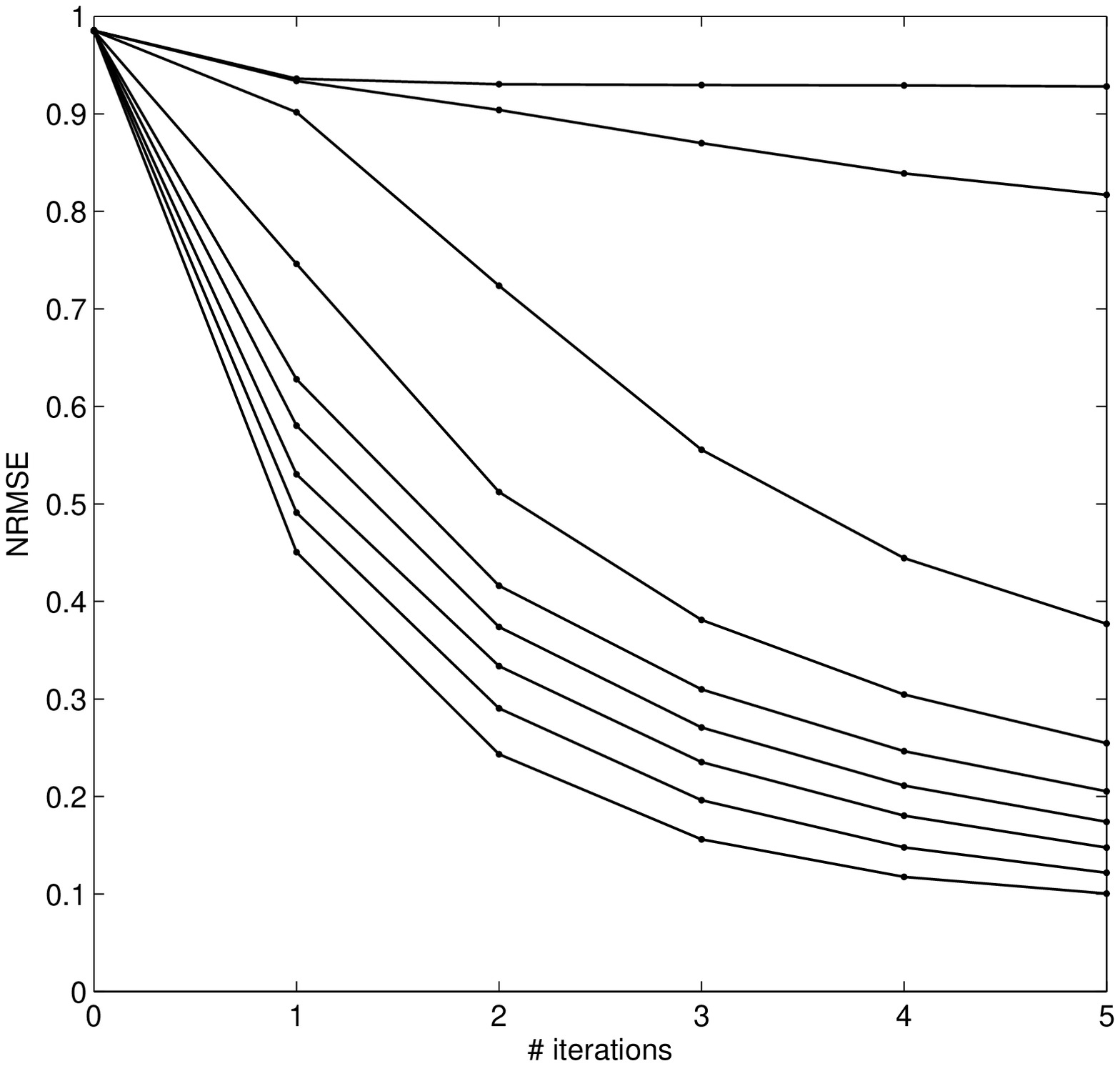}\\
  (c)\includegraphics[width=6cm]{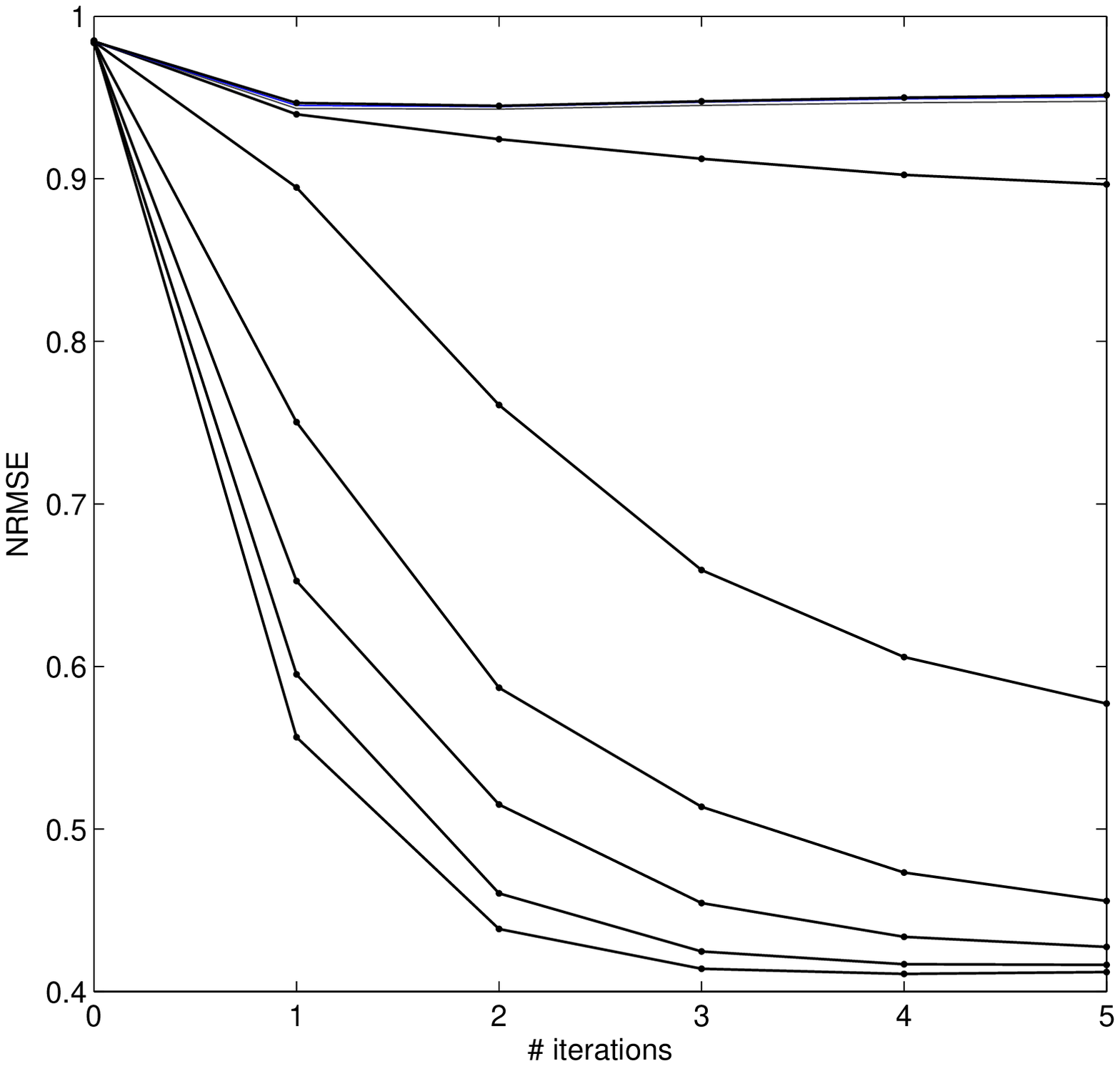}
  (d)\includegraphics[width=6cm]{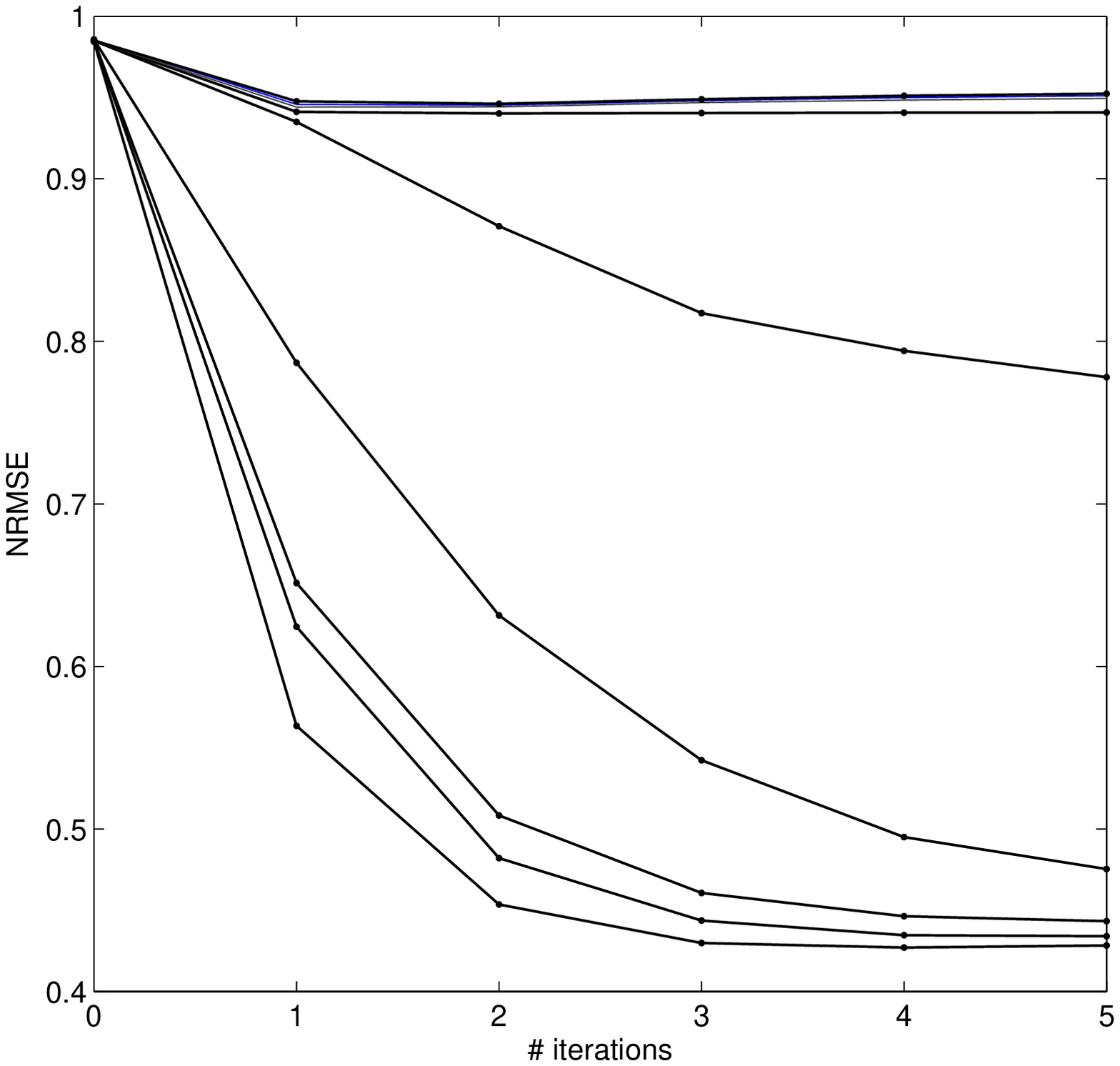}\\
  \caption{Prediction error evolvement of Bayesian spectral method for $N=100$, $D=30$, $M=10$.
  (a) linear matching, (b) thresholded ($f(x)=\tanh(x)$) match function ($\beta=1$)(c) noisy linear matching, ,
(d) noisy thresholded matching.}\label{fig:1}
\end{figure}

Using $N$, $D$, $M$ as free parameters, we randomly generated two matrices $\mathbf{X}$
and $\mathbf{W}$ with each components to be either 1 or -1. The match matrix $\mathbf{Y}$ was then
computed as in section 2. We then investigated the capabilities
of our algorithm in reconstructing the full matrix for 4 different cases: nearly perfect linear matching
(little noise with $\sigma=0.01$ added), noisy linear matching ($\sigma=0.1$), nearly perfect thresholded match
function ($f(x)=\tanh(x)$), and noisy thresholded matching.

Fixing $N=100$ and $D=30$, we applied the algorithm to two different data sets of  $M=5$ and $M=10$.
For each case, the unknown positions of the data matrix were randomly picked with missing percentages
from 10\% to 90\%. For each missing percentage, an average result from 1000 runs were then
obtained. The accuracy of predictions was measured by the popular used normalised
root mean square error (NRMSE) after each iteration.

Figure~\ref{fig:1} shows the evolvement of the algorithm prediction error through the first 5
iterations. The error at iteration 0 was calculated by replacing each missing position by
the mean value of its corresponding subject. The 9 lines correspond to 9 different missing
percentages, ranging from 10\% to 90\%. It can be seen that the algorithm converge quickly to
reasonable accuracy after 5 iterations for up to 40\% missing percentages in all cases.
With too much missing data (80\% upward for perfect linear data), the matrix could not be
reconstructed. This phase transition corresponds to a \emph{rigidity}
percolation threshold. The noise moves this threshold down to 60\%, but under perfect data
collection condition, the nonlinear match function ($f(x)=\tanh(x)$) does not make any considerable
effect. It comes to action however in the case of noisy data, moving the percolation threshold
to 50\%. In the case of linear matching data with very little noise, the rigid percolation
threshold actually agrees with the theoretical result in ~\cite{maslov01}, which equals to
$p=1-2M/N\simeq 80\%$

\begin{figure}
  \centering
  (a)\includegraphics[width=6cm]{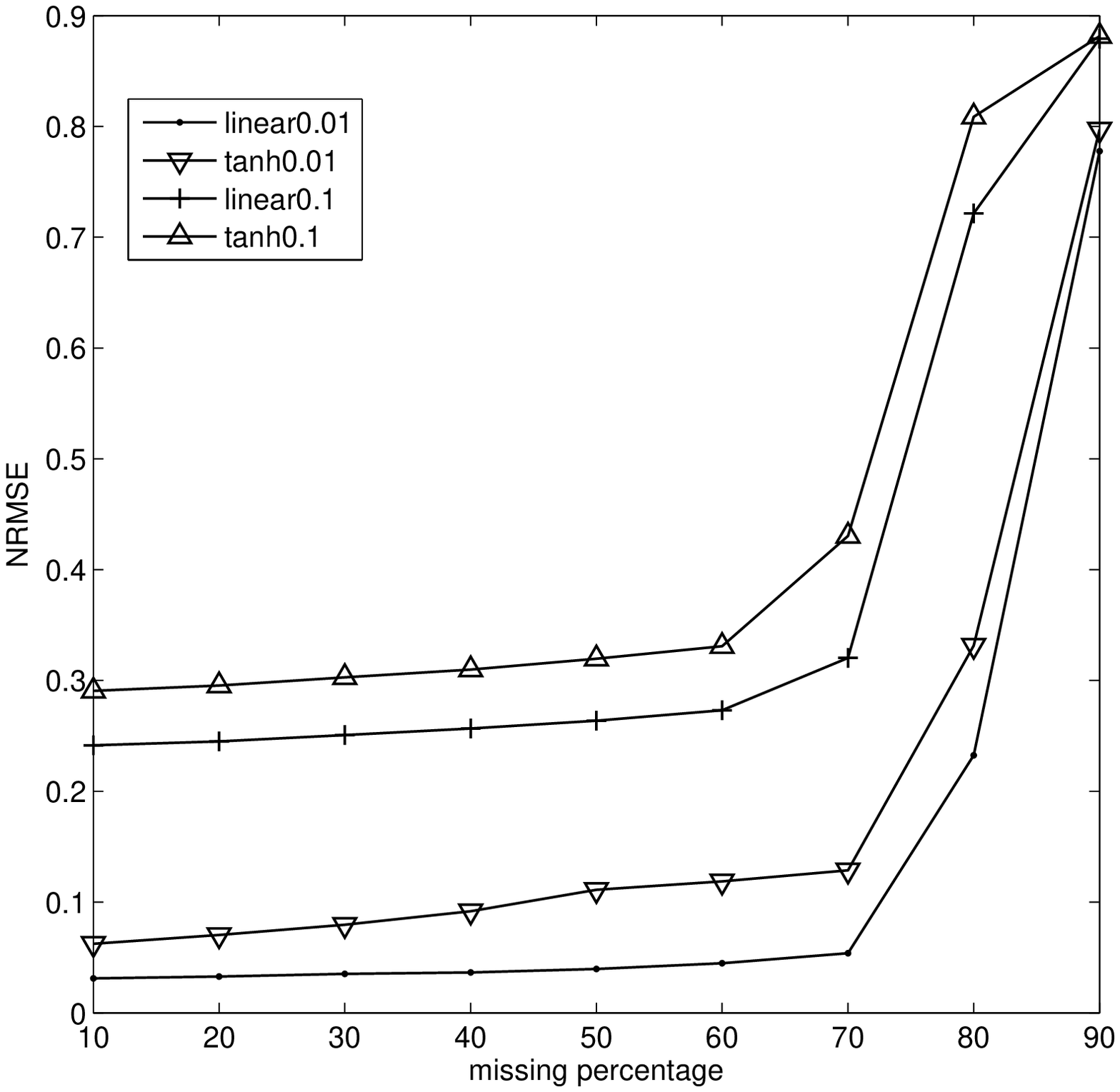}
  (b)\includegraphics[width=6cm]{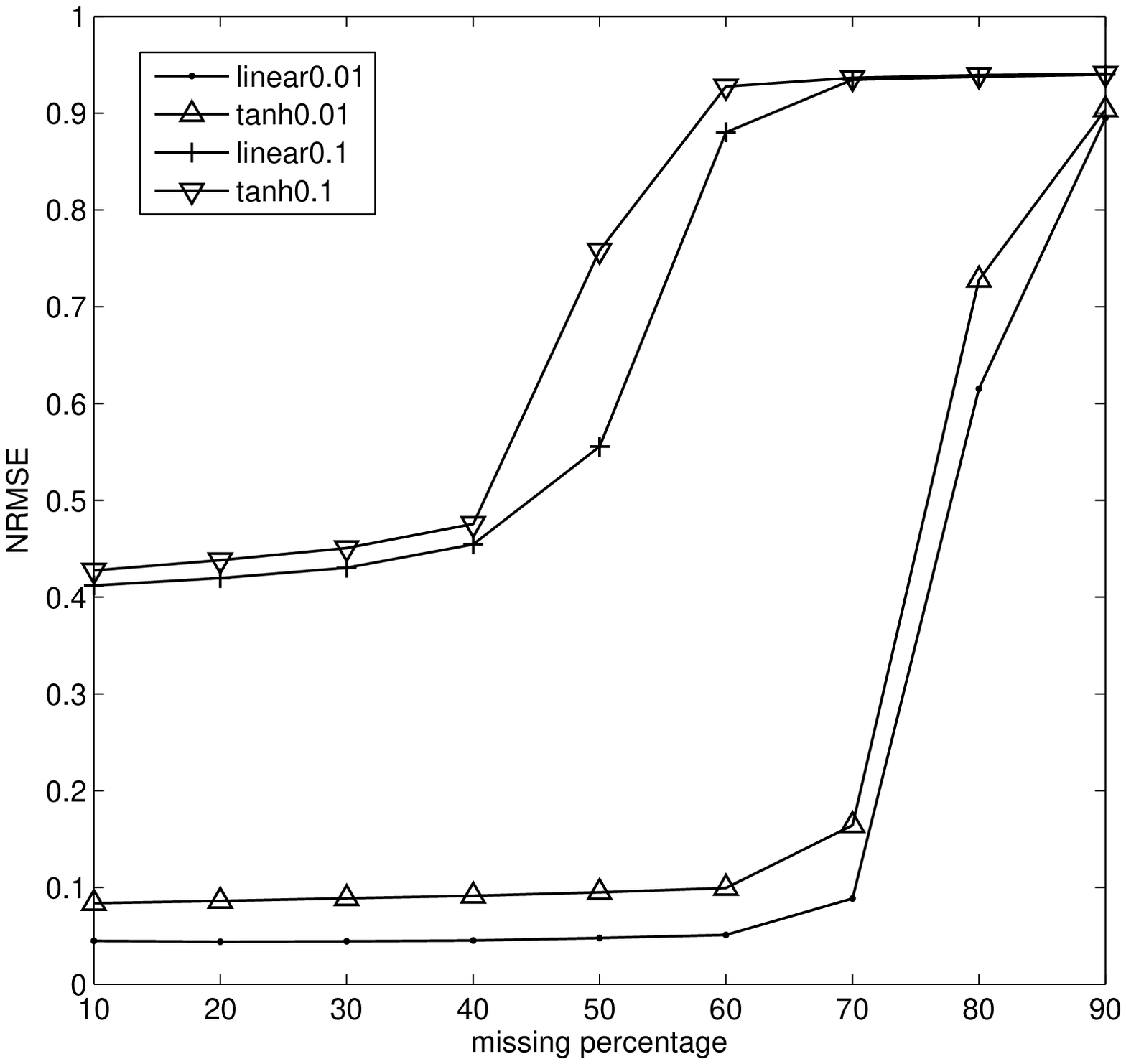}\\
  \caption{Prediction errors of Bayesian spectral method for the 4 reference cases. (a)$M=5$, (b)$M=10$}\label{fig:2}
\end{figure}

Figure~\ref{fig:2} presents the effect of noise and nonlinearity on prediction accuracy. It
could be easily seen the dramatic increase of error at the corresponding percolation thresholds.
Noise shows critical effect on prediction accuracy compared to the thresholded match function,
increasing by the complexity of data (large $M$).

\subsection{Real data}

To get a reasonable view of the phase transition, we applied the Bayesian spectral method on two different data sets. The first one is the measurements of the transcription
levels under different experimental conditions of 215 mutants of essential yeast genes \cite{Mnaineh}. The dataset was extracted by the authors from microarray measurements of about 5000 genes of the budding yeast S.cerevisiae.
The second dataset is the small-scale measurements of binding energies between the bacteria E.coli ligands and enzymes of the bacteria E.coli \cite{macch04}. We removed all columns with missing values from the data, leaving two complete data matrices of size $215 \times 15$ and $119 \times 15$.

Following similar procedure to synthetic data tests, we randomly generated a list of missing positions in the complete data matrix, and cleared the known values out of those cells. We then applied the Bayesian spectral method to reconstruct the missing values on each data set, and
compared to the original ones to calculate the prediction error. The procedure was applied for various missing percentage from 5\% to 90\%, for each case the average result from 1000 runs was obtained.

\begin{figure}[t]
  \centering
  (a)\includegraphics[width=6cm]{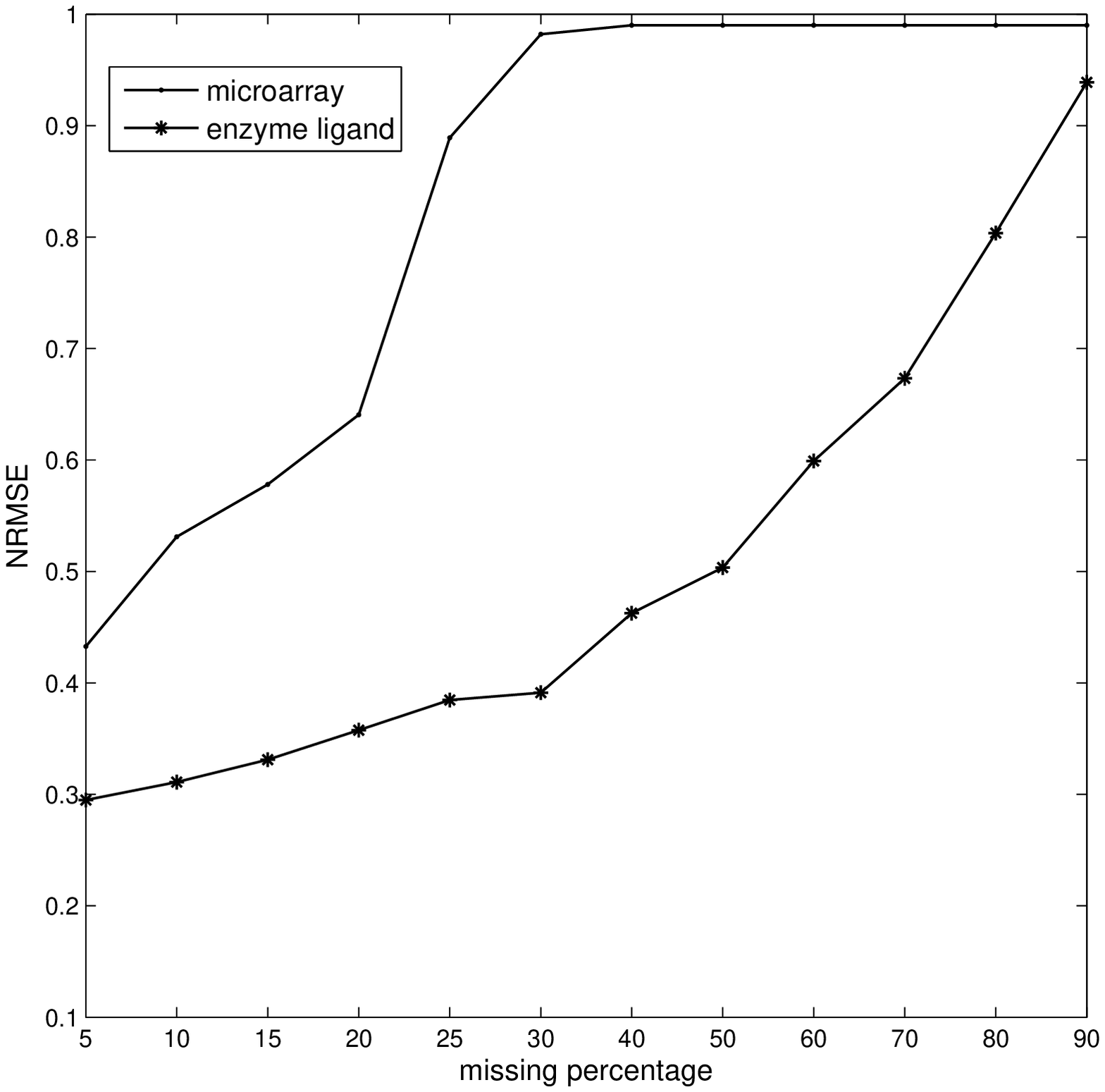}
  (b)\includegraphics[width=6cm]{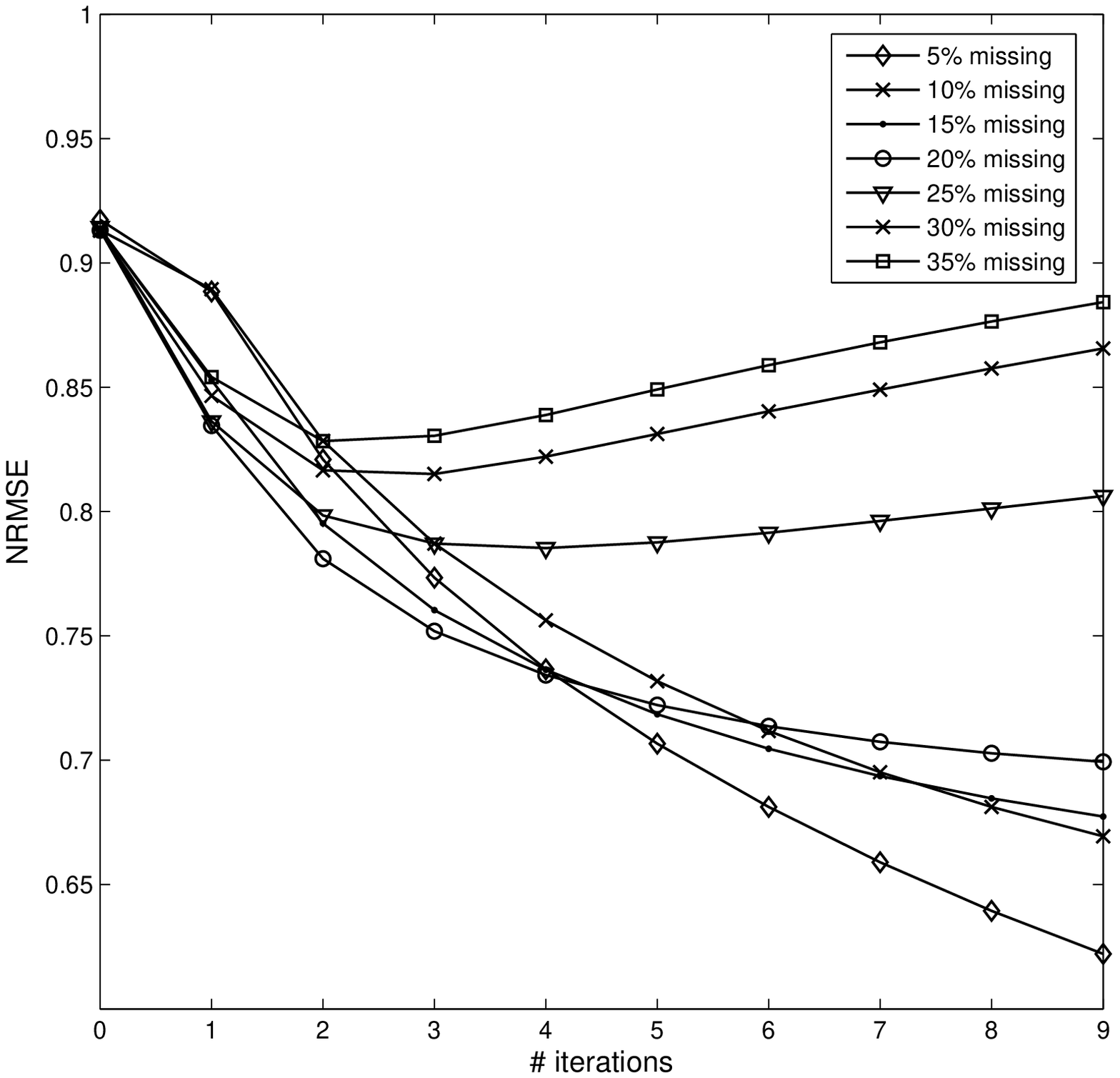}\\ 
  \caption{Missing value reconstruction of microarrays by the Bayesian spectral method. (a)prediction error for increasing missing percentage, (b)error evolvement of microarray data }\label{fig:3}
\end{figure}

The prediction results over increasing missing percentages of the two datasets are shown in figure \ref{fig:3}a.
The algorithm performs considerably better on the enzyme-ligand data than microarray data. Setting a threshold at a value of the NRMSE of $0.7$, where there is a big jump for the microarray data, we see that the algorithm is able to reconstruct enzyme-ligand values up to 70\% missing, while the transition occurs at 20\% missing for the case of microarray. Since enzyme-ligand data set was acquired by focused biochemistry studies of each enzyme, its reliability is a lot higher compared to the high-thoughput technique used to obtain gene expressions \cite{drag06}. Apart from the common inaccuracy due to background noises and diversity of samples, the limitation of the measuring technology  largely reduces microarray data reliability.

Figure~\ref{fig:3}b shows the prediction evolvement though early iterations for microarray data. Agreeing with the final prediction results, the algorithm kept improving its prediction by iterations with up to 20\% missing data, while not much information can be reconstructed for data with 25\% upward missing. 
Interestingly, in this case the error diminishes during the first 2 iterations, but then increases again.
A similar effect was observed~\cite{unpublished} in reconstructing missing values in synthetic data by means of linear regression and multi-entry correlations. Using the simplest algorithm~\cite{bagnoli04,DiPatti}, missing values $\tilde{y}_{ij}$ are reconstructed by using two-entry correlations: $\tilde{y}_{ij} \propto \sum_k C_{jk} y_{ki}$, where $C_{ij}$ is the Pearson correlation among entries of subjects $i$ and $j$. For a large percentage of missing values, the correlation is affected by a low statistics, so one may try to exploit three- and higher-entry correlations. According with the level of noise, the number of hidden components and the nonlinearity of the matching function, this procedure improves the results only up to a certain point, after which the multi-entry contributions essentially furnish more noise than information, in a way similar to what is observed in Figure~\ref{fig:3}b. By looking at Figure~\ref{fig:1}, one can realize that in the case of noisy match (subfigures c and d), the evolvement curves of the error above the phase boundary (not converging phase) slightly rise after the first two iterations, while in the absence of noise (subfigures a and b) it keeps decreasing. This behavior suggests that in case of microarray, the level of noise is much greater that the level of nonlinearity.

\section{Conclusions}


We have applied a Bayesian spectral algorithm in order to investigate the nature of noise in synthetic and real datasets. The investigation on synthetic data shows that the approach is very robust in handling large percentage of missing data both in terms of accurate prediction and quick convergence rate. Although a solid conclusion on the ability of our method in predicting systems noise nature has not been made, the result on synthetic data and on some experimental datasets are very promising. A systematic investigation with specific consideration on different kinds of noises would prove useful.

\end{document}